\documentclass[a4paper,11pt]{article}
\usepackage{jinstpub} 
\usepackage{siunitx}
\usepackage{booktabs}
\usepackage{overpic}
\usepackage{tikz}

\usetikzlibrary{arrows.meta, positioning, shapes.geometric, fit}
\newlength{\ul}
\title{\boldmath The BUTTON-30 detector at Boulby}

\author[\,]{\large BUTTON Collaboration:\\[2mm]}
\author[a]{J.~Bae,}
\author[b]{M.~Bergevin,}
\author[b]{E.~P.~Bernard,}
\author[c]{D.~S.~Bhattacharya,}
\author[d]{J.~Boissevain,}
\author[e]{S.~Boyd,}
\author[f]{K.~Bridges,}
\author[g]{L.~Capponi,}
\author[f]{J.~Coleman,}
\author[h]{D.~Costanzo,}
\author[i]{T.~Cunniffe,}
\author[b]{S.~A.~Dazeley,}
\author[j]{M.~V.~Diwan,}
\author[b]{S.~R.~Durham,}
\author[c]{E.~Ellingwood,}
\author[k]{A.~Enqvist,}
\author[h]{T.~Gamble,}
\author[j]{S.~Gokhale,}
\author[f]{J.~Gooding,}
\author[a]{C.~Graham,}
\author[k]{E.~Gunger,}
\author[m]{J.~J.~Hecla, \footnote{Present address: Department of Nuclear Science and Engineering, Massachusetts Institute of Technology, 77 Massachusetts Ave. Cambridge, MA, 02139, USA}}
\author[l]{W.~Hopkins,}
\author[a]{I.~Jovanovic,}
\author[m,n]{T.~Kaptanoglu,}
\author[h]{E.~Kneale,}
\author[m,n]{L.~Lebanowski,}
\author[i]{K.~Lester,}
\author[b]{V.~A.~Li,}
\author[h]{M.~Malek,\footnote{Present email address: \href{mailto:matthew@banterbots.ai}{matthew@banterbots.ai}}}
\author[d]{C.~Mauger,}
\author[f]{N.~McCauley,}
\author[f]{C.~Metelko,}
\author[g]{R.~Mills,}
\author[f]{A.~Morgan,}
\author[c]{F.~Muheim,}
\author[i]{A.~Murphy,}
\author[c]{M.~Needham,}
\author[a]{K.~Ogren,}
\author[m,n]{G.~D.~Orebi~Gann,}
\author[i]{S.~M.~Paling,}
\author[o]{A.~F.~Papatyi,}
\author[p]{A.~Petts,
\footnote{visiting professor at: Department of Physics, Durham University, DH1~3LE, UK}}
\author[i]{G.~Pinkney,}
\author[i]{J.~Puputti,}
\author[q]{S.~Quillin,}
\author[e]{B.~Richards,}
\author[j]{R.~Rosero,}
\author[h]{A.~Scarff,\footnote{Present address: TUBR \url{http://www.tubr.tech}}}
\author[f]{Y.~Schnellbach,\footnote{Present address: Science for Nuclear Diplomacy, Technische Universität Darmstadt, Schlossgartenstr.~7, 64289 Darmstadt, Germany}}
\author[i]{P.~R.~Scovell,}
\author[l]{B.~Seitz,}
\author[h]{L.~Sexton,}
\author[c]{O.~Shea,}
\author[c]{G.D.~Smith,\footnote{Present Address: Health Physics, Gartnavel Royal Hospital, 1055 Great Western Road, Glasgow~G12~0XH, UK}}
\author[r]{R.~Svoboda,}
\author[e]{D.~Swinnock,}
\author[f]{A.~Tarrant,}
\author[l]{F.~Thomson,}
\author[f]{J.~N.~Tinsley,}
\author[i]{C.~Toth,}
\author[s]{M.~Vagins,}
\author[j]{G.~Yang,}
\author[j]{M.~Yeh,}
\author[e]{E.~Zhemchugov}

\affiliation[a]{Department of Nuclear Engineering and Radiological Sciences, University of Michigan, Ann Arbor, MI~48109, USA}
\affiliation[b]{Lawrence Livermore National Laboratory, Livermore, CA~94550, USA}
\affiliation[c]{School of Physics and Astronomy, University of Edinburgh, Edinburgh, EH9~3FD, UK}
\affiliation[d]{Department of Physics and Astronomy, University of Pennsylvania, Philadelphia, PA~19104, USA}
\affiliation[e]{Department of Physics, University of Warwick, Gibbet Hill Road, Coventry, CV4~7AL, UK}
\affiliation[f]{Department of Physics, University of Liverpool, Merseyside, L69~7ZE, UK}
\affiliation[g]{United Kingdom National Nuclear Laboratory, Cumbria, CA14~3YQ, UK}
\affiliation[h]{School of Mathematical and Physical Sciences, University of Sheffield, Sheffield, S10~2TN, UK}
\affiliation[i]{STFC, Boulby Underground Laboratory, Boulby Mine, Redcar-and-Cleveland, TS13~4UZ, UK}
\affiliation[j]{Brookhaven National Laboratory, Upton, NY~11973, USA}
\affiliation[k]{Department of Materials Science \& Engineering, Nuclear Engineering Program, University of Florida, Gainesville, FL~32611-6400, USA}
\affiliation[l]{School of Physics and Astronomy, University of Glasgow, Glasgow, G12~8QQ, UK}
\affiliation[m]{University of California, Berkeley, Berkeley, CA, USA}
\affiliation[n]{Lawrence Berkeley National Laboratory, Berkeley, CA, USA}
\affiliation[o]{Pacific Northwest National Laboratory, 902 Battelle Boulevard, Richland, WA~99352, USA}
\affiliation[p]{EDF Energy UK, Hartlepool, TS25~2BZ, UK}
\affiliation[q]{AWE Aldermaston, Reading, Berkshire, RG7~4PR, UK}
\affiliation[r]{University of California at Davis, Department of Physics and Astronomy, Davis, CA~95616, USA}
\affiliation[s]{Department of Physics and Astronomy, University of California, Irvine, Irvine, CA~92697-4575, USA}

\emailAdd{mneedham@ed.ac.uk}

\abstract{The BUTTON-30 detector is a 30-tonne technology demonstrator designed to evaluate the potential of hybrid event detection, simultaneously exploiting both Cherenkov and scintillation light to detect particles produced in neutrino interactions. The detector is installed at a depth of 1.1 km in the Boulby Underground Laboratory allowing to test the performance of this new technology underground in a low background environment. This paper describes the design and construction of the experiment.}


\begin{document}
\maketitle
\flushbottom
\newcommand{\BUTTON}{BUTTON\mbox{-}30~}
\section{Introduction}
\label{sec:intro}
The study of the properties and behaviour of neutrinos provides a means to probe fundamental physics, to understand fusion processes within the Sun, and enables remote monitoring of nuclear reactors \cite{SajjadAthar:2021prg, Akindele:2021sbh}. Detecting neutrinos remains one of the most demanding frontiers in particle physics, requiring large detection volumes, sophisticated sensor technologies, and media capable of capturing the signatures of interactions. The detection of neutrinos with energies of $\sim 3$ MeV is particularly challenging because of the low interaction cross-section and large backgrounds from environmental radioactivity and spallation products from cosmic rays. Typically, low-energy neutrino interactions are reconstructed using either water Cherenkov detectors or oil-based scintillator detectors. However, in the last decade a worldwide program has begun to develop hybrid detectors that take advantage of both Cherenkov and scintillation light in one technology \cite{Aberle:2013jba,Anderson:2022lbb,Bat:2021jyq,Guo:2017nnr,Land:2020oiz,Zhao:2023ydx, ANNIE:2017nng,Ascencio-Sosa_2024,Theia:2019non}. These studies have identified water-based liquid scintillator (WbLS) with gadolinium (Gd) doping as a potentially powerful new technology for future neutrino experiments. The addition of a water-based scintillator increases the light yield while seeking to preserve the directionally-sensitive topology of the prompt Cherenkov photons. Gadolinium doping captures the neutrons produced by neutrinos interacting by inverse $\beta$-decay, allowing tagging the interaction by the subsequent gamma-ray cascade. In this paper, the construction and installation of the Boulby Underground Technology Testbed for Observing Neutrinos, a 30 tonne hybrid optical detector (BUTTON-30) that aims to test this technology is described.  The host site for BUTTON-30 is the Boulby Underground Laboratory (BUL), the UK’s deep underground science laboratory located in an active polyhalite mine in north-east England.

Located at 1.1\,km depth, BUL \cite{Murphy:2012zz} is one of the few underground laboratories around the world that is suitable for experiments requiring low background radiation conditions. The overburden of \SI{2.8}{\km} water equivalent \cite{Robinson:2003zj} attenuates the cosmic radiation flux by a factor of $10^6$.  In addition, the laboratory has low background radiation due to the minimal radionuclide content of the surrounding salt rock: $\sim\!0.1$ ppm of uranium and thorium, and 1,130 ppm of potassium \cite{Malczewski:2013lqy}. Radon levels are also low, averaging $2.4\,\textrm{Bq}/\textrm{m}^3$ \cite{Scovell:2023wnk,Scovell_2024}. \BUTTON is the largest detector constructed in the Boulby mine to date.

\BUTTON is developed in collaboration with complementary efforts elsewhere, including the EOS experiment at Lawrence Berkeley National Laboratory \cite{Anderson:2022lbb} and the WbLS development program at Brookhaven National Laboratory \cite{Xiang:2024jfp}. While those experiments focus on surface-level deployments, \BUTTON provides an opportunity to benchmark the technology in a deep underground, low-background environment. This allows the characterisation of detector performance including scintillation properties, neutron backgrounds, and the response of novel photodetectors such as the LAPPD \cite{LAPPD:2016yng} or wavelength shifting plates, in conditions relevant to large-scale subterranean neutrino observatories. \BUTTON will mitigate risks and provide a validation platform for future kilotonne-scale detectors. The demonstrator is designed to:
\begin{itemize}
\item	Validate Gd-WbLS stability, performance, and purification in an underground setting;
\item	Benchmark advanced photodetector technologies;
\item	Develop and test calibration systems, DAQ integration, and clean material handling under realistic deployment constraints;
\item	Establish engineering and operational procedures for underground deployment of Gd-WbLS.
\end{itemize}	
By achieving these goals, \BUTTON will significantly reduce the technical and programmatic risks associated with scaling to kilotonne detectors. The experiment will also allow studies of the atmospheric neutrino flux and cosmogenic neutron production in the environment of BUL. In this way, the project will contribute to the international roadmap toward hybrid Cherenkov-scintillator neutrino observatories and supports cross-cutting applications in nonproliferation, nuclear safeguards, and fundamental science. This paper is arranged as follows. First, the materials used in the detector are discussed. In the following sections, the layout of the tank and optical detection and calibration system, the data acquisition, and expected rates from radioactive backgrounds are described.%

\section{Material Selection and Radioassay}
\label{sec:clean}
Detector operation with Gd-loading or WbLS requires careful material selection to prevent leaching and corrosion into the fill medium \cite{Xiang:2024jfp, MARTI2020163549}. Impurities in the water would degrade the optical transparency of the water and hence impact the detector performance. Only marine grade 316 stainless steel is used for metal parts that are in contact with the fill medium. To further improve resistance to corrosion, the surface of all metallic parts was passivated using nitric acid (pickling). In addition to this treatment, the mounting frames and metallic parts of the optical modules were electropolished. Only plastic materials that showed no evidence of leaching in the WbLS soak tests were selected for use. In particular, Viton$^{\textrm{TM}}$ is selected for O-rings and gaskets.

To quantify background rates in the detector, radioassays of all detector materials were made prior to installation. All samples were tested using the suite of HPGe detectors hosted by the BUGS (Boulby UnderGround Screening) facility~\cite{SCOVELL2018160,Scovell_2024}. Additional glass sample assays were performed at UC Davis and LBNL. The results of this test program are detailed in Table~\ref{tab:backgrounds}. Most of the samples have radioactivity levels below the sensitivity limits of the assay systems used. These results feed into the background model for BUTTON described in Section~\ref{sec:sim}.

To ensure that the detector components were not contaminated during transport, each item was double-bagged. Before entering the Class 10000 cleanroom space of the laboratory, the outer bag was removed and the component was placed in temporary storage. Immediately prior to installation, the components were removed from their inner bag.

\begin{table}[htb!]
\centering
\caption{Summary of radioassay results for major materials used in the BUTTON-30 detector construction. Listed activities are given per unit mass and scaled by component mass to obtain estimated total contributions. Additional measured isotopes (not shown) include $^{60}$Co and $^{54}$Mn, while $^{235}$U activities are inferred from the activity ratio A($^{235}$U):A($^{238}$U) = 1:21.7.}
\label{tab:backgrounds}
\begin{tabular}{c c c c c c c c}
\toprule
& \textbf{Estimated} & \multicolumn{3}{c}{\textbf{Assay results [mBq/kg]}} & \multicolumn{3}{c}{\textbf{Estimated Activity [Bq]}} \\
\textbf{Component} & \textbf{mass [kg]} & \textbf{$^{238}$U} & \textbf{$^{232}$Th} & \textbf{$^{40}$K} & \textbf{$^{238}$U} & \textbf{$^{232}$Th} & \textbf{$^{40}$K}\\
\midrule
\textbf{Tank} &&&&&&& \\
Steel & 10000 & <3.2 & <2.5 & <9.7 & <32.0 & <25.0 & <97.0 \\ \\
\textbf{PSUP} &&&&&&& \\
Steel box tubes & 680 & <0.7 & 0.5$\pm$0.2 & <3.1 & <0.48 & 0.34 & <2.11 \\
Steel cylinders & 320 & 19$\pm$2 & <0.7 & <8.8 & 6.08 & <0.22 & <2.82 \\ \\
\textbf{Liner} &&&&&&& \\
Polyethylene & 3.2 & 903$\pm$10 & 58$\pm$4 & 1516$\pm$66 & 2.89 & 0.19 & 4.85 \\
Zipties & 1.5 & <14 & <8.7 & <153 & <0.02 & <0.01 & <0.23\\ \\
\textbf{PMT} &96 PMTs& \multicolumn{3}{c}{Assay results from one PMT} &\multicolumn{3}{c}{Activity for 96 PMTs} \\
Borosilicate glass & 96 & 1.62 [Bq] & 1.05 [Bq] & 0.11 [Bq] & 155 & 101 & 10.7 \\ \\
\textbf{Encapsulation} &&&&&&& \\
UV acrylic dome & 165 & 70$\pm$4 & <2.9 & 61$\pm$44 & 11.5 & <0.48 & <10.0 \\
Acrylic dome & 197 & <7.7 & <3.9 & <34 & <1.51 & <0.77 & <6.69 \\
Steel plates & 370 & <20 & <4.0 & <50 & <7.42 & <1.48 & <18.5\\
M8 bolts & 66 & <16 & 21$\pm$6 & <45 & <1.21 & 1.59 & <34.1 \\
Optical gel & 114 & <6.9 & <3.5 & <46 & <0.79 & <0.40 & <5.24 \\
Coax cable & 9.2 & 234$\pm$19 & 70$\pm$11 & 720$\pm$152 & 2.14 & 0.64 & 6.59 \\
Silica desiccant & 10.0 & 198$\pm$4 & 450$\pm$6 & 670$\pm$35 & 1.99 & 4.51 & 6.72 \\
\bottomrule
\end{tabular}
\end{table}

\section{Tank design}
\label{sec:tank}
\begin{figure}[htb!]
    \centering
    \includegraphics[width=0.50\linewidth]{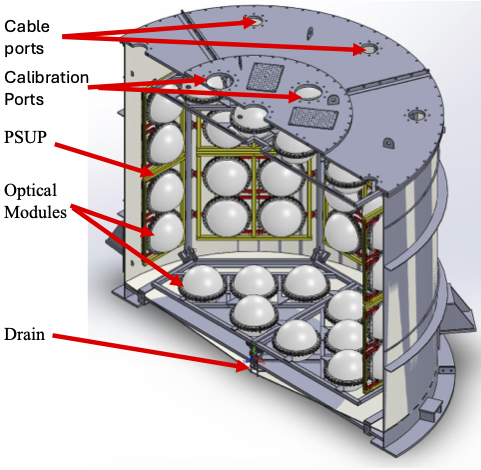}
     \includegraphics[width=0.44\linewidth]{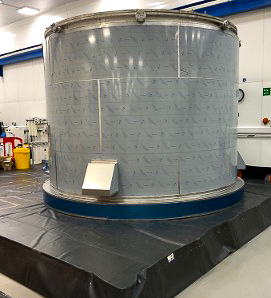}
    \caption{(Left) CAD model of the BUTTON-30 detector tank, illustrating the placement of the PMT support structure (PSUP) and the mounting positions of the optical modules. (Right) Photograph of the fully constructed tank installed in the Boulby Underground Laboratory.}
    \label{fig:tank}
\end{figure}
The detector is housed in a right cylindrical tank with (Figure~\ref{fig:tank}) an outer diameter of \SI{3.6}{\metre} and a height of \SI{3.2}{\metre} constructed from 8\,mm thick marine grade 316 stainless steel.  This tank size gives a detector volume large enough for technology demonstration purposes while meeting the constraints imposed by the current laboratory size. Tank assembly was carried out in the underground laboratory space using non-thoriated continuous seam welding. Contamination from the welding process was removed with a Tungsten Inert Gas (TIG) brush. The welded seams were tested with dye penetrant testing to ensure that the tank is watertight and capable of holding a static liquid load of 30,000\,$\mathrm{kg}$. The tank lid has ports for cable routing and to allow the deployment of radioactive sources for calibration. The outside of the tank is insulated with Armaflex$^{TM}$ to allow operation at a water temperature of \SI{\sim 10}{\celsius}.

\begin{figure}[htb!]
    \centering
    \includegraphics[width=0.99\linewidth]{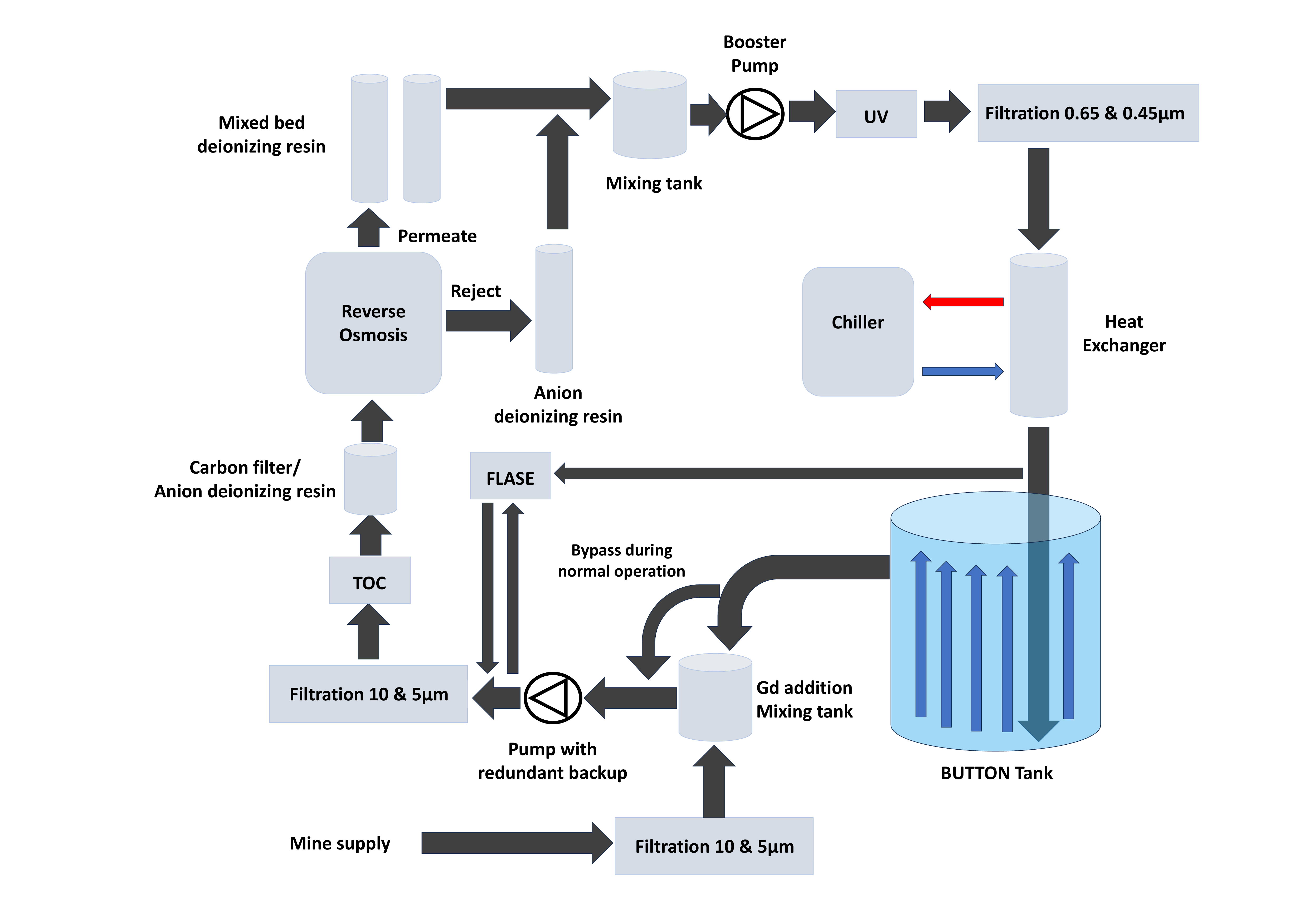}
    \caption{Schematic of the BUTTON-30 water purification and circulation system. The diagram shows the filtration, deionization, and bacterial removal stages used to maintain high optical transparency and compatibility with both gadolinium-doped water and WbLS.}
    \label{fig:filter}
\end{figure}

Good optical transparency is crucial to ensure efficient light collection by the optical detection system (Section~\ref{sec:ods}).  This requires the use of Type 1 grade laboratory water with a resistivity of \SI{18}{\mega\ohm\centi\metre}
and 1--5 parts per billion total organic carbon (TOC) content and a water filtration system (Figure~\ref{fig:filter}) designed to remove bacteria and material impurities. The filtration system, profiting from developments for previous experiments \cite{MARTI2020163549,Fischer:2020hjj,Zhao:2023ydx},  is  compatible with both Gd doping and WbLS operation.

The system is designed to run at 21\,L/min giving an approximate full replacement of the tank volume once per day. Water from the tank first passes through a set of standard filters and the TOC lamps. The subsequent reverse osmosis (RO) system removes much finer impurities, with up to 99\% contaminant removal down to a size of 0.001~$\mu$m. A mixed bed deionizing resin system allows targeted removal of different ions in the fill media. Finally, UV lamps kill and destroy bacteria before the water is further filtered and returned to the tank via a heat exchanger. Although all the system parts have been confirmed to be WbLS compatible, a different configuration of UV lamps and RO after deployment WbLS will be required. This is due to the potential damage UV light and passage through the RO may cause to the scintillator portion of the WbLS. Hence, the WbLS will need to be separated into LS and water before the RO and the LS purified in a secondary system. This configuration also mitigates any risk that WbLS will damage the RO membranes.

\section{Optical detection system}
\label{sec:ods}
\begin{figure}[htb!]
    \centering
    \includegraphics[width=0.55\linewidth]{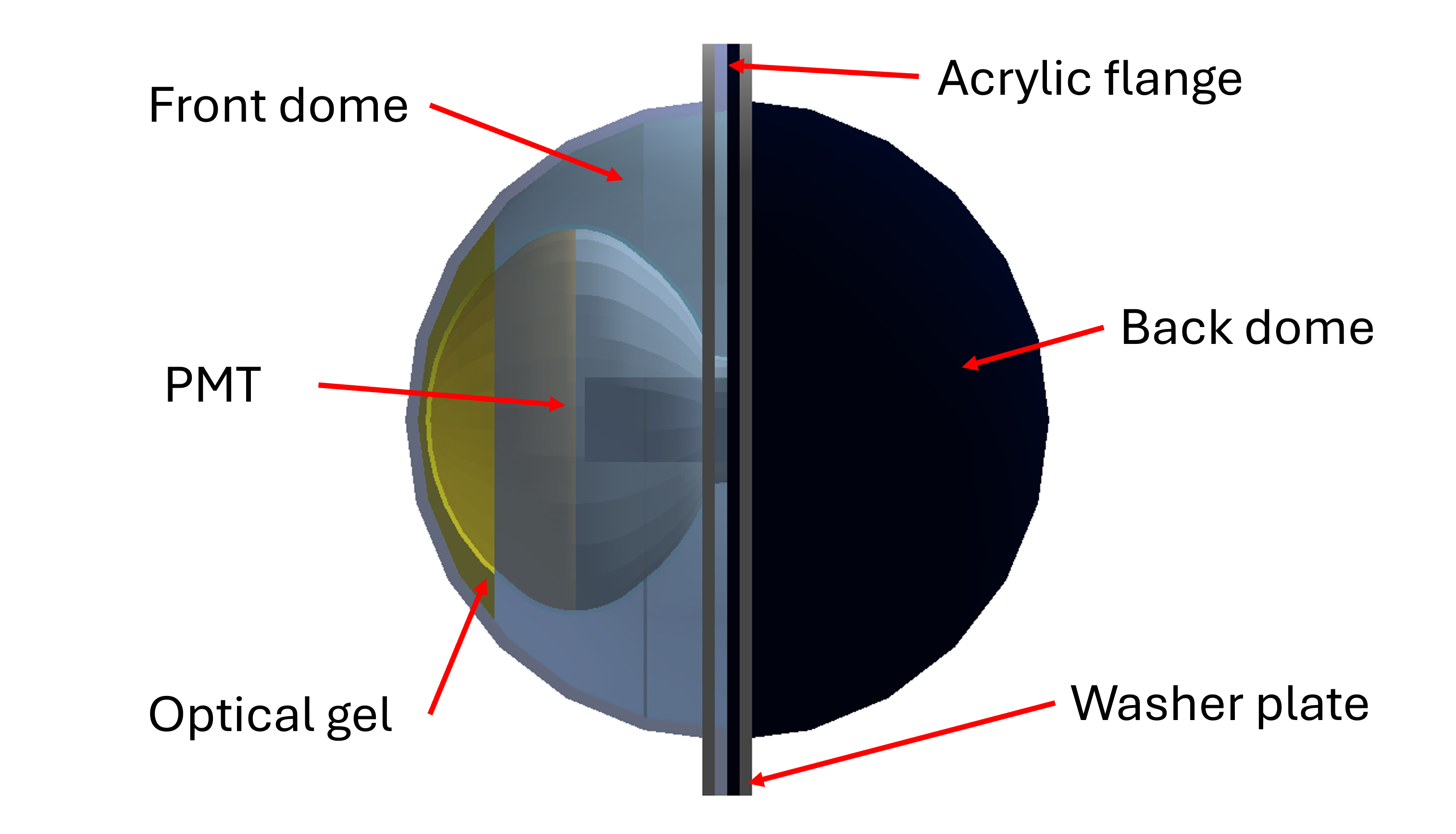}
    \caption{Geant4 simulation model of an encapsulated optical module.}
    \label{fig:optical}
\end{figure}
The optical detector system consists of ninety-six 10-inch Hamamatsu R7081-100 photomultiplier tubes (PMTs) \cite{R7081datasheet} with low-radioactivity glass bulbs encapsulated in acrylic housings. The PMTs have a quantum efficiency of around $25 \, \%$ dark count rates in the range $1$--$8\,$kHz and give $10^7$ gain at operating voltages in the range $1$--$2\,\textrm{kV}$. More details on the electrical characterisation of a subset of these PMTs are described in~\cite{Akindele:2023ixz}. Though the PMT bases are qualified for use in Gd-doped water, they are known from soak tests to be incompatible with operation in WbLS. To allow data collection with a variety of fill media, the PMTs are encapsulated in acrylic housings (Figure~\ref{fig:optical}) to form watertight optical modules. A paper documenting the production of the housings is in preparation \cite{housingpaper}. Each housing consists of two (roughly) hemispherical blow moulded acrylic domes with a diameter of $48.6\,$cm and a flat flange. The thickness of the acrylic varies across the dome from around $3\,$mm at the apex to $6\,$mm  at the flange. The chosen size is the minimum required to fit the PMT base and to route out the cable without excessive bending. The front half of the housing uses acrylic that is transparent to wavelengths in the UV, whilst standard acrylic (with the internal surface painted black) is used for the back half. RTV27905 silicone gel is used to provide an optical coupling between the acrylic and the PMT glass. Simulation studies using as input the measured transmittance of the acrylic and gel show that encapsulation reduces the optical transmission by $12\,$\% compared to a bare PMT. The two halves are sealed with a Viton O-ring and stainless steel washer plates to provide compression. High-voltage (HV) and signal return is provided by an integrated cable attached to the PMT base, which exits the housing via a water-tight penetrator piece. All the housing material in contact with the water was chosen based on compatibility with both Gd loading and WbLS operation \cite{Xiang:2024jfp, MARTI2020163549}. In addition, a prototype housing was operated submerged in $1\,\%$ WbLS for several weeks with no visible sign of degradation.

\begin{figure}[htb!]
    \centering
    \includegraphics[width=0.44\linewidth]{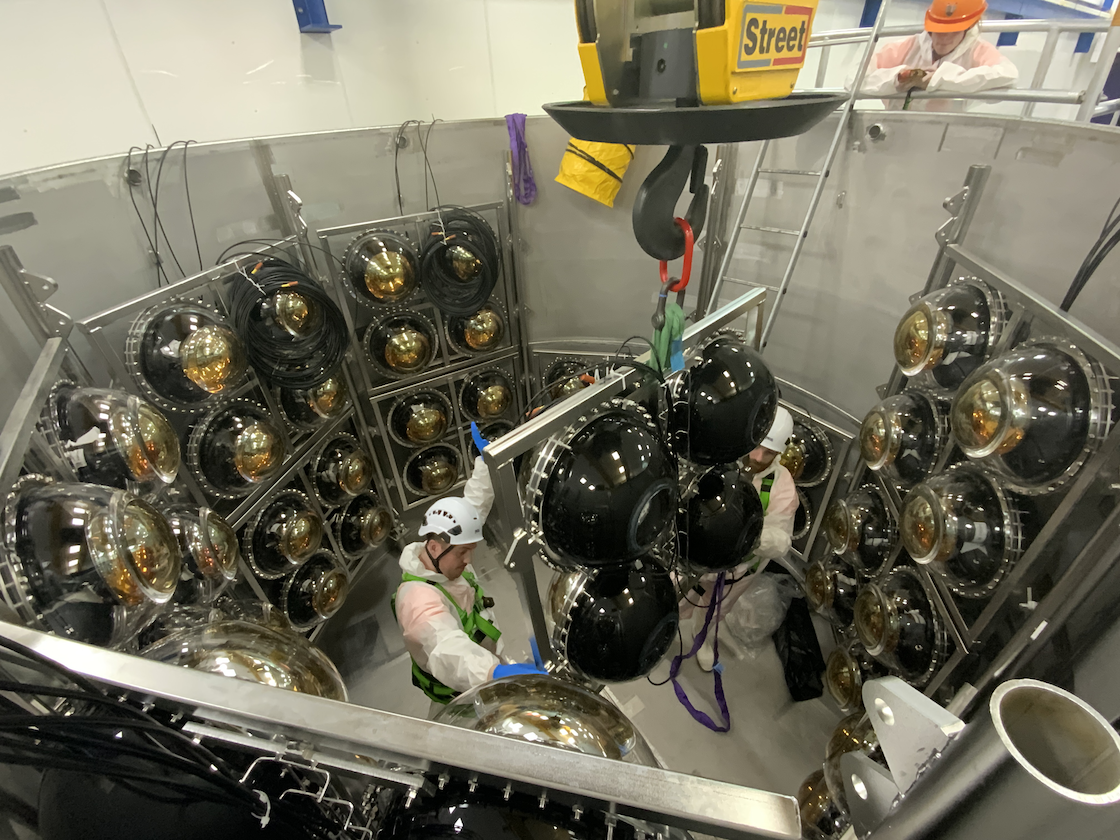}
    \includegraphics[width=0.48\linewidth]{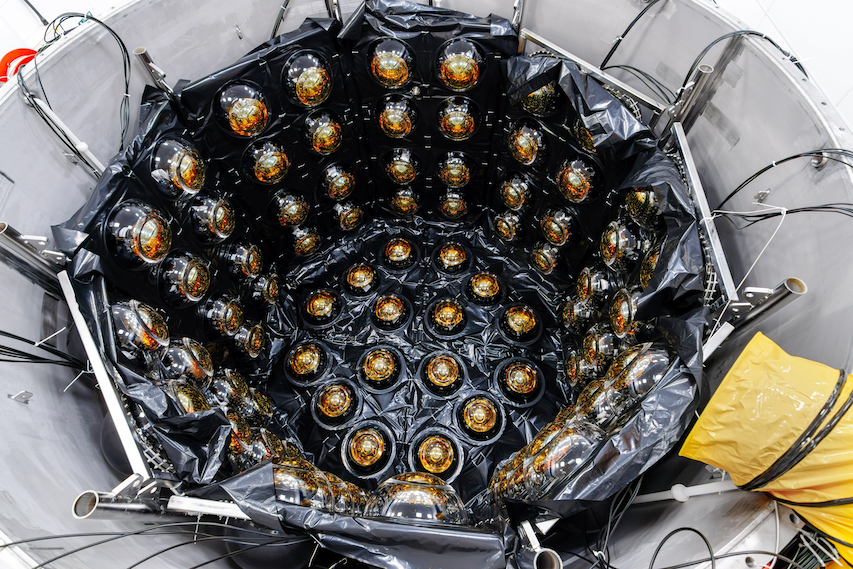}
    \caption{(Left) Installation of the barrel optical modules on the PSUP. (Right) Photograph of the optical modules mounted on the PSUP barrel frames inside the tank, showing the arrangement used to achieve uniform photocoverage. The black polyethylene liner is also visible.}
    \label{fig:psup}
\end{figure}

The encapsulated PMTs are mounted inside the tank on a stainless steel support structure referred to as the PSUP (Figure~\ref{fig:psup}).  Sixty-four PMTs are arranged radially into an octagonal barrel, with a further sixteen mounted on each of the top and the bottom frames. To maximize the possible fiducial volume within the small \BUTTON tank, the PSUP is mounted as closed to the tank wall as possible. Consequently, allowing for the housing size the PMT photocathode sits around $50\,\textrm{cm}$ from the tank wall giving an active water volume of around $12\,\textrm{m}^3$. The photocoverage at the liner surface is $\sim13\,\%$. There is space on the top frames to allow future deployment of additional photodetectors such as LAPPDs \cite{LAPPD:2016yng,ANNIE:2025vra}, wavelength shifting plates \cite{Mullen:2022nwg} or calibration PMTs. The PSUP frames were welded with nonthoriated rods and then pickled and passivated prior to transport to Boulby. Before installation in the tank, each frame was visually inspected. Any part showing evidence of corrosion was cleaned with a TIG brush. 

To provide an internal optical barrier and to reduce the impact of light scattered from the surfaces of the PSUP and optical modules, a black polyethylene liner is mounted on to the PSUP structure. Polyethylene was chosen from several tested candidates because of its availability and resistance to leaching in Gd-doped water and WbLS, on the basis of UV-visibility tests. 

\section{Calibration}
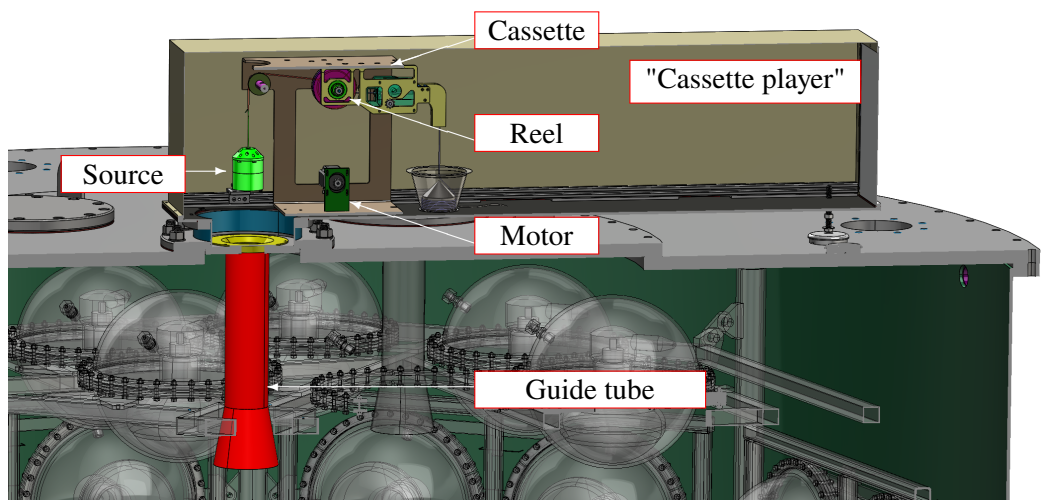
\begin{figure}[htb!]
\centering
   \begin{overpic}[width=0.9\linewidth]{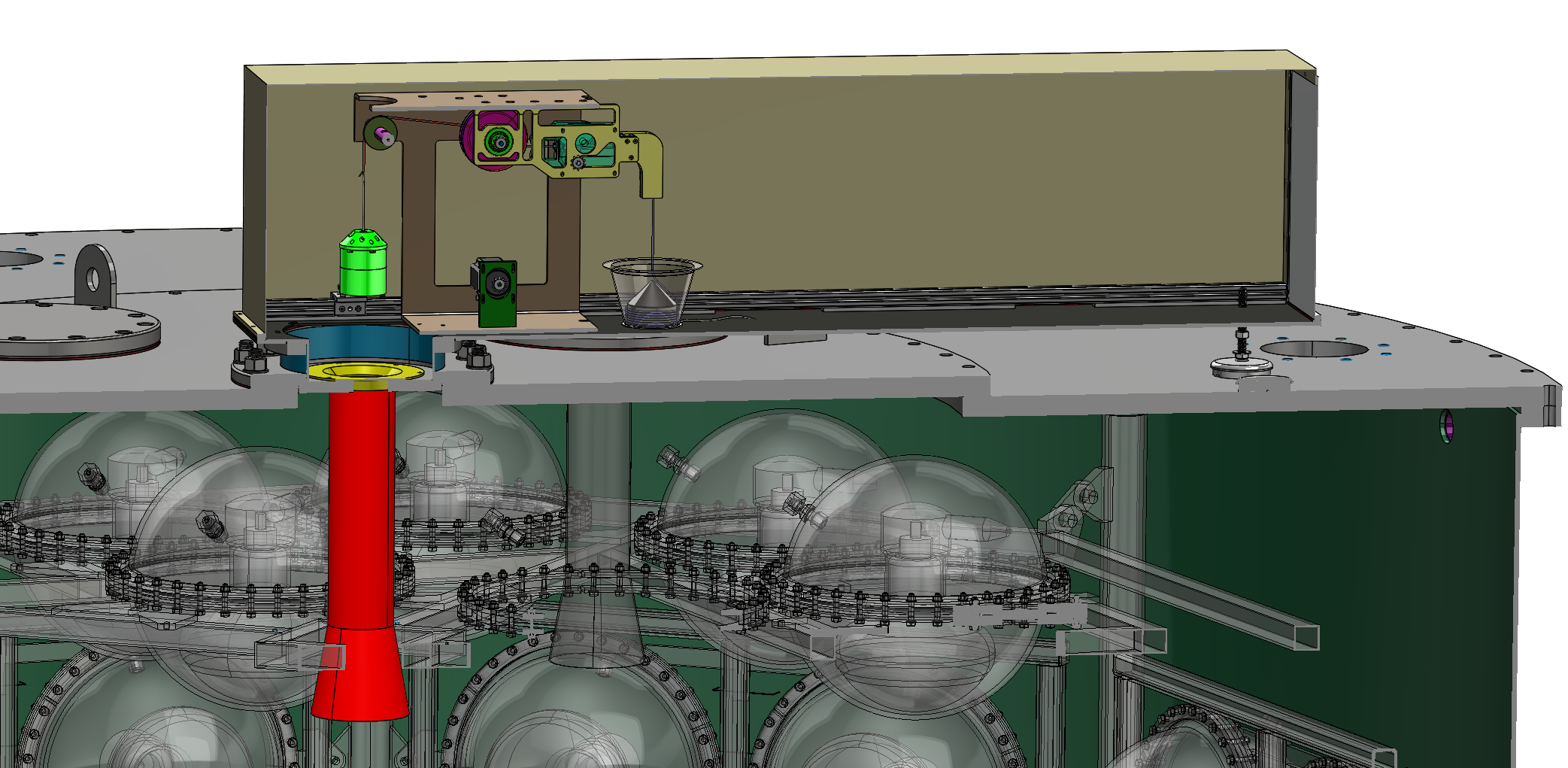}
   \put(10,32){\color{white}\vector(1,0){11}}
   \put(5,31){\fcolorbox{red}{white!30}{\parbox{40pt}{\centering \textcolor{black}Source}}}   
   \put(45,45){\color{white}\vector(-3,-1){8}}
   \put(45,45){\fcolorbox{red}{white!30}{\parbox{40pt}{\centering \textcolor{black}Cassette}}}  
   \put(45,35){\color{white}\vector(-3,1){12}}
   \put(45,35){\fcolorbox{red}{white!30}{\parbox{40pt}{\centering \textcolor{black}Reel}}}   
   \put(45,25){\color{white}\vector(-3,1){12}}
   \put(45,25){\fcolorbox{red}{white!30}{\parbox{40pt}{\centering \textcolor{black}Motor}}}   
   \put(45,11){\color{white}\vector(-1,0){20}}
   \put(45,10){\fcolorbox{red}{white!30}{\parbox{80pt}{\centering \textcolor{black}Guide tube}}}  
   \put(60,40){\fcolorbox{red}{white!30}{\parbox{80pt}{\centering \textcolor{black}"Cassette player"}}}   
   \end{overpic}
\caption{CAD model of the calibration source cassette, including the reel, motor, sealed source container, and the cassette deployment system mounted on the tank lid for positioning radioactive calibration sources.
\label{fig:cassette}}
\end{figure}

Radioactive sources, a light injection system and cosmic muons will be used to calibrate the PMT response. The radioactive sources will be deployed from housings mounted on the tank lid, shown in Fig.~\ref{fig:cassette}. These allow the installation of cassettes consisting of a sealed source container, a motor, and a reel. Two types of source containers will be utilised. The first is relatively small and designed to deploy a radioactive check source. Initially, an untagged Americium-Beryllium (AmBe) source  producing a 4.4-MeV gamma-ray and a correlated single neutron will be deployed.  The untagged 100-$\mu$Ci ($\sim$$3.7$-$\mathrm{MBq}$) source is housed in an electropolished 316L stainless steel capsule with thick walls that suppress low-energy x rays; water tightness is provided by a O-ring seated in a machined groove. The main stainless body is $7\ \mathrm{cm}$ in diameter with an internal cavity of approximately $2.6\ \mathrm{cm}$ by $2.6\ \mathrm{cm}$; the lid is secured using six $7.6\ \mathrm{cm}$ long 10-24 stainless socket head cap screws engaging PTFE-coated stainless steel nuts retained in slots. 

The tagged AmBe assembly comprises a 1-inch Hamamatsu R1924A-100 high-QE PMT optically coupled to a 550-g BGO crystal in a sealed container. BGO provides an efficient detection of the 4.4-MeV gamma-ray and a compact footprint for tagging. The PMT operates with positive high voltage; a single coaxial umbilical carries both HV and signal to an external splitter similar to those used for the main \BUTTON PMTs. The mechanical package is an elongated cylinder with a conical cap, similar to the untagged source envelope for consistent deployment through the lid housings. A dedicated cassette and reel manage the HV coax during lowering and retrieval, maintaining strain relief and clean routing into the detector. The BGO detector allows the generation of an electronic tag  when a 4.4-MeV gamma-ray emitted by the AmBe source interacts with the BGO crystal, indicating the simultaneous emission of a neutron. In the future, additional sources will be deployed.

The light-injection system allows for continuous PMT monitoring during data collection. It consists of four diffusers mounted on the PSUP. The design of the diffusers is similar to those planned to be used in the HyperKamiokande experiment \cite{Hyper-Kamiokande:2018ofw}. Each diffuser illuminates the PMTs with a range of light wavelengths to track the single photoelectron response of each PMT over time. The diffuser is designed to deliver a broad cone of light over a half-angular range of 40 degrees (similar to a Cherenkov cone). It consists of a PTFE hemisphere with a radius of $1.5\,$cm contained within a watertight housing. Light is transported from an external light source to the diffusers via four multimode, step-index Thorlabs$^{\textrm{TM}}$ FG105UCA armoured optical fibres. The laser system consists of a PicoQuant laser diode producing short (50 ps) pulses of 405 nm light. This system can be upgraded for additional wavelengths by adding new diode modules. The laser light is distributed to the four diffusers and a monitoring photomultiplier using a set of commercial optical switches. The light source system is integrated into the data acquisition system so that it can be operated from the surface control room. From simulation studies, it is expected that the system will be able to monitor the single photoelectron peak with an uncertainty of around 5\,\% and allow timing offsets between PMTs to be determined with a precision of $0.5 \, \textrm{ns}$. In addition to the wall-mounted diffusers, a small diffuser ball is envisioned to be deployable at regular intervals through the central calibration cassette player. This will be a small PTFE sphere connected to the laser system, similar to that deployed in the ANNIE experiment \cite{ANNIE:2015inw}, and will provide intermittent timing calibration points with a precision of $1 \, \textrm{ns}$. 

\begin{figure}[t!]
\centering
\includegraphics[width=1\linewidth]{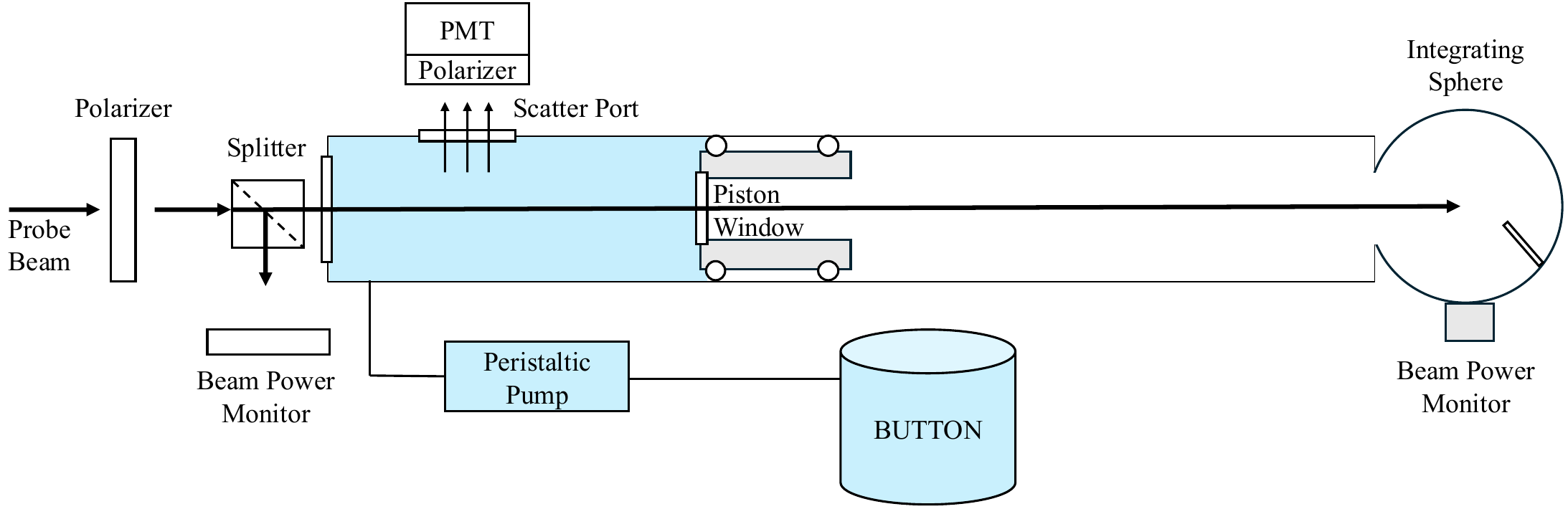}
    \caption{A schematic of the proposed FLASE attenuation and scattering device. Figure adapted from the LASE experiment~\cite{Hecla:2023rrp}}
    \label{fig:FLASE_dia}
\end{figure}

\begin{figure}[t!]
\centering
\includegraphics[width=0.99\linewidth]{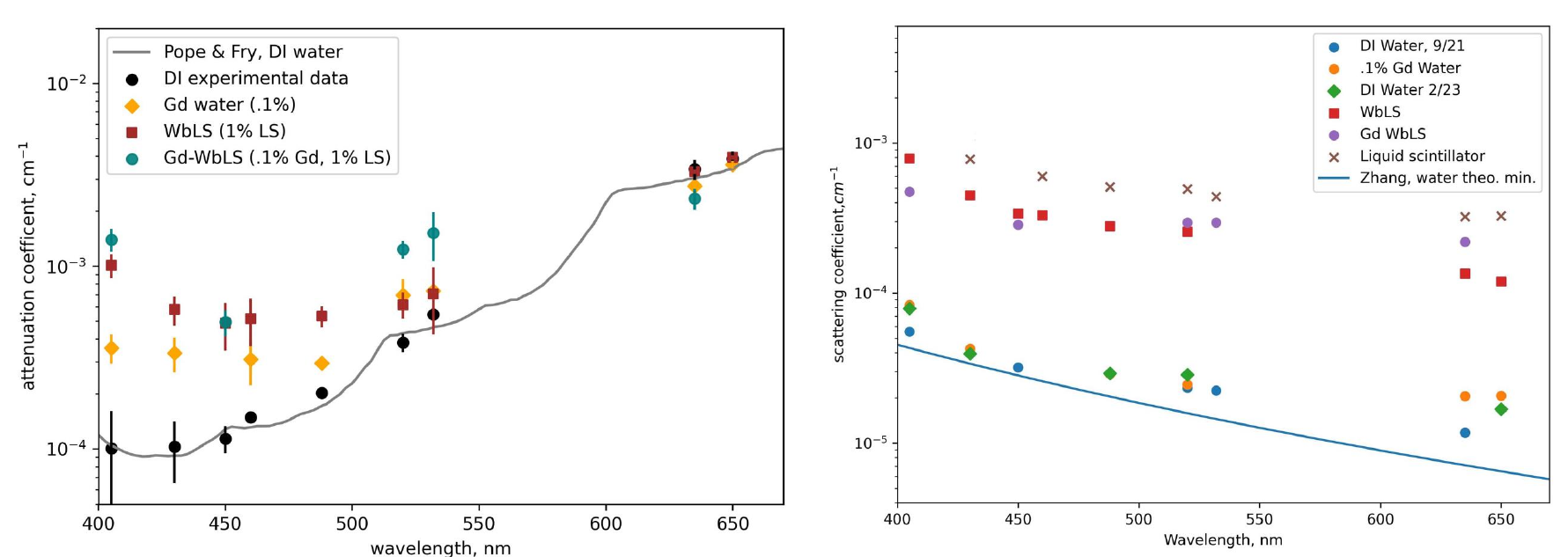}
    \caption{(Left) Attenuation measurements taken by the LASE experiment for different fills: pure de-ionized water, water doped with 0.1~\% Gd, $1~\%$ WbLS, and $1~\%$ WbLS plus $0.1~\%$ Gd. The solid black line is from \cite{Pope:97}.  (Right) Scattering measurements for the same media. The solid blue line is from \cite{Zhang:09}. Both figures taken from ~\cite{Hecla:2025arz}. }
    \label{fig:FLASE_meas}
\end{figure}

A liquid diagnostic tool, called FLASE (Florida-Livermore Attenuation and Scattering Experiment) will be constructed, similar to the Livermore Attenuation and Scattering Experiment (LASE) system described in~\cite{Hecla:2023rrp}, and is intended for direct measurements of photon attenuation and scattering in liquids. The instrument will be deployed horizontally and adjacent to the \BUTTON experiment to enable periodic direct measurements of liquid from the detector. The liquid will be accessible directly from both before and after the liquid purification system. A schematic of the proposed device and previous measurements by the LASE experiment are shown in Figures~\ref{fig:FLASE_dia} and \ref{fig:FLASE_meas}. 


Attenuation in the fill media will be quantified by monitoring laser power as a function of distance through a horizontal liquid column. 
The instrument uses a movable piston to control the length of the liquid column. 
The piston moves by the displacement of liquid into and out of the column using a peristaltic pump, which changes the length of the column of liquid. 

The liquid column and the piston are contained within a $2.8\,\mathrm{m}$ stainless steel tube, $5.1\,\mathrm{cm}$ in diameter and mounted horizontally on an optical table for vibration stability. 
By adjusting the position of the piston/window, the liquid column length can be adjusted anywhere within a range from $0.4$ to $2.8\,\mathrm{m}$. 
The initial set of laser wavelengths used are 430, 460, 488 and 505 nm. 
Laser beam power is measured before it enters the liquid column using a photodetector. 
The beam then enters the liquid through a window and travels through a set of baffles to minimize back-scattering. 
A window is incorporated into the piston design so that laser light exits the liquid and travels to a photodetector mounted inside an integrating sphere at the end of the instrument to measure the power of the exiting laser beam as a function of the length of the liquid column.

The liquid scattering length is measured using a PMT housed 90 degrees from the liquid column near the start of laser path through the liquid column, shown in Fig.~\ref{fig:FLASE_dia}. A rotatable polarizing screen is placed in the input laser path to control input laser polarization. An additional rotatable screen is placed at the exit of the scattering port and immediately before the PMT. The two screens can be used to diagnose the type of scattering present, since the Mie and Rayleigh scattering components are sensitive to the polarization angle.  
Scattering-length measurements taken with the prototype LASE are included in Fig.~\ref{fig:FLASE_meas}.
The uniqueness of the system lies in its compactness, the ability to simultaneously measure both scattering and attenuation properties, and the capability to measure light transport as a function of liquid column length, by adjusting the position of the piston within the tube using the peristaltic pump.

\section{DAQ and data processing}
\label{sec:daq}
\begin{figure}[htb!]
\centering
\begin{tikzpicture}[
  node distance=0.7\ul and 0.7\ul,
  every node/.style={draw, minimum height=0.25\ul, minimum width=0.7\ul},
  arrow/.style={-{Stealth}, thick},
  ]
\node[ellipse] (PMT) {PMT};
\node[right=of PMT] (Splitter) {Splitter};
\node[right=0.75\ul of Splitter, yshift=0.25\ul] (ADC) {ADC};
\node[right=0.75\ul of Splitter, yshift=-0.25\ul] (HV) {HV};

\node[draw=none, fit=(ADC) (HV), inner sep=20pt] (ADC_HV) {};

\node[right=of ADC_HV, anchor=center, yshift=0pt] (RBU) {RBU};
\draw[arrow] (PMT) -- (Splitter);
\draw[arrow] (Splitter) -- (PMT);
\draw[arrow] (Splitter) -- (ADC);
\draw[arrow] (HV) -- (Splitter);
\draw[arrow] (ADC) -- (RBU);
\draw[arrow] (ADC) -- node[above, fill=none, draw=none] {data} (RBU);
\draw[arrow] (RBU) -- node[below, fill=none, draw=none] {control} (HV);
\end{tikzpicture}
\caption{Block diagram of the BUTTON-30 readout chain. The diagram traces the flow of signals from the photomultiplier tubes (PMTs) through the splitter boards and digitizers to the Readout Buffer Unit (RBU), including trigger logic and data handling steps.}
\label{fig:readoutchain}
\end{figure}
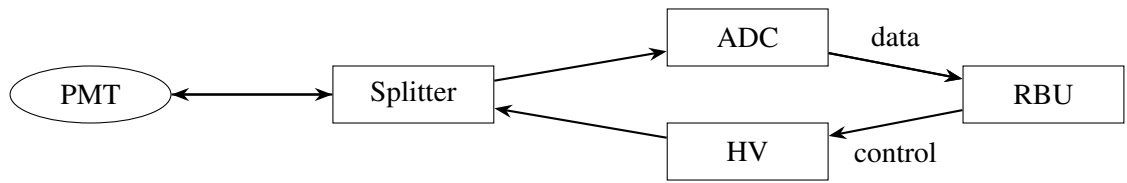

\begin{figure}[b!]
    \centering   \includegraphics[width=0.9\linewidth]{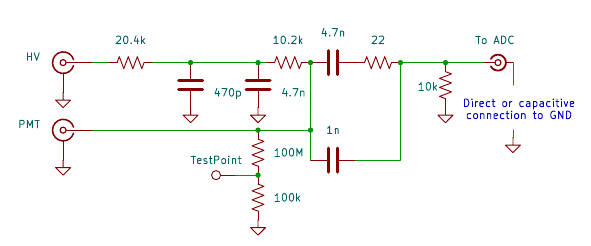}
    \caption{Electrical layout of the HV splitter.}
    \label{fig:splitter}
\end{figure}

The layout of the readout chain is shown in Figure~\ref{fig:readoutchain}. The PMTs are supported by a total of 112 analog-to-digital converter (ADC) channels. This enables full detector readout and support for auxiliary subsystems such as the calibration system. Each PMT has a single coaxial cable HV supply and signal return. The signal and HV are separated using custom-designed splitter boards (Fig.~\ref{fig:splitter}). Bias for the PMTs is provided by 16 CAEN V6534 VME-based HV supplies, each with six independent channels.  The HV boards are managed via a CAEN V1718 VME master module and connected to the main Readout Buffer Unit (RBU-01) via USB. 

\begin{figure}[htb!]
    \centering   \includegraphics[width=0.8\linewidth]{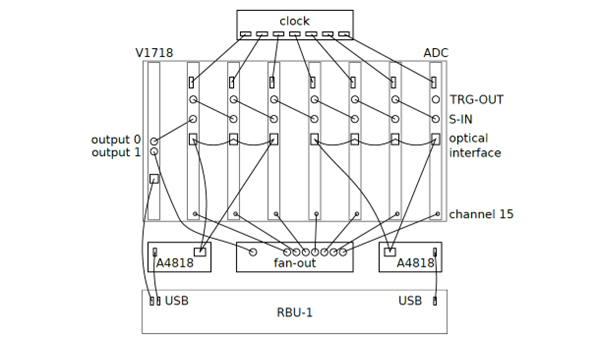}
    \caption{Layout of the ADC crate connections for BUTTON-30. The figure shows the routing of digitized PMT signals, distribution of the common 62.5 MHz clock, and clock-reset synchronisation used to align timing across all digitizer modules.}
    \label{fig:adcsync}
\end{figure}

Signal digitization is performed by seven CAEN V1730 series digitizers. Each module provides 16 input channels and operates at $500\,\mathrm{MS/s}$ (sampling every 2~ns) with digital pulse processing capabilities. These ADCs feature waveform capture and sub-nanosecond precision trigger time measurements. Synchronisation across the digitizers is achieved using a CAEN DT4700 clock generator, which supplies a common 62.5-MHz clock signal (16-ns period) to all ADCs. To synchronise their internal clocks, a reset signal is sent to all boards. Due to inconsistent signal propagation delays of up to $\sim\!16\,\mathrm{ns}$, a calibration procedure is needed (Figure~\ref{fig:adcsync}). This involves issuing a software acquisition start pulse followed by a series of 100 calibration pulses. These are fanned out using a CAEN N625 quad linear fan-in/fan-out unit, connected to a second CAEN V1718 master module in the ADC crate. This allows for the calibration of the relative time offsets between the ADC channels to a precision of $100\,\textrm{ps}$.

DAQ operations are orchestrated by a single multi-core 2U readout server located underground and running ToolDAQ~\cite{richards:ToolDAQ, Richards:2019lkb}. This collects digitized data, applies trigger logic, sorting, and filters events. The software framework is responsible for the full DAQ chain, including multi-threaded data processing, slow-control and monitoring interfaces.

Upon detection of a pulse, the corresponding channel sends a data packet to the readout PC. This consists of pulse processing data and includes a $300\,\mathrm{ns}$-wide waveform of 150 time samples. Since the electronics operate asynchronously, the packets arrive unordered on the PC. The RBU collates these data and groups them into $100\,\mathrm{ms}$  time slices sorted on the basis of the timestamp to recover the correct sequence of events.

Multiple triggers exist in the system for different data sources. The baseline is a majority trigger that fires if the number of hits in a dynamic sliding window exceeds a set threshold. The size of the window is set to $100\,\mathrm{ns}$ which is broad enough to allow for the $\sim 20\,\textrm{ns}$ spread of photon arrival times for Cherenkov light produced in the tank (1) and the $16\,\textrm{ns}$ synchronisation spread of the ADC channels. The default threshold is four hit PMTs which is the minimum needed to reconstruct a vertex. With a trigger rate of $\sim\!120\, \textrm{Hz}$ (Section~\ref{sec:sim}) around 2.5 Terabyte of data will be collected per day.  Triggers for calibration and a zero-bias trigger are also implemented. The zero-bias trigger allows the readout of random time windows to study backgrounds and PMT pedestals.

\section{Radioactive background rates}
\label{sec:sim}
\begin{table}[hp!]
\centering
\caption{Calculated activities and trigger rates for dominant decay chains for the rock of the cavern, tank, and liner. Trigger rates are shown for individual isotopes and decay products. The last column shows the rate for $>3$ PMTs registering a hit in a 100~ns time window. This is the default trigger condition. 
}
\label{tab:backgroundtriggerstank}
\begin{tabular}{c c c c c c c}
\toprule
& & & \textbf{Activity} & \textbf{Detection} & \textbf{Detection rate}\\
\textbf{Component} & \textbf{Decay chain} & \textbf{Isotope} & \textbf{rate [Bq]} & \textbf{rate [Hz]} & \textbf{(hits>3) [Hz]} \\
\midrule
\textbf{Rock} & $^{238}$U  & $^{234}$Pa & 2.19 $\times$ $10^{5}$ & 0     & 0        \\
     & $^{238}$U  & $^{214}$Pb & 2.19 $\times$ $10^{5}$ & 0.788 & 0        \\
     & $^{238}$U  & $^{214}$Bi & 2.19 $\times$ $10^{5}$ & 555   & 44.0     \\
     & $^{238}$U  & $^{210}$Tl & 43.8 & 0.179 & 1.73 $\times$ $10^{-2}$ \\
     & $^{238}$U  & $^{210}$Bi & 2.19 $\times$ $10^{5}$ & 0     & 0        \\
     & $^{232}$Th & $^{228}$Ac & 3.38 $\times$ $10^{4}$ & 15.7  & 3.38 $\times$ $10^{-2}$ \\
     & $^{232}$Th & $^{212}$Pb & 3.38 $\times$ $10^{4}$ & 0     & 0        \\
     & $^{232}$Th & $^{212}$Bi & 2.16 $\times$ $10^{4}$ & 2.62  & 0.118    \\
     & $^{232}$Th & $^{208}$Tl & 1.22 $\times$ $10^{4}$ & 124  & 23.6     \\
     & $^{40}$K   & $^{40}$K   & 1.05 $\times$ $10^{3}$ & 0.313 & 1.36 $\times$ $10^{-2}$\\ \\
     
\textbf{Tank} & $^{238}$U & $^{234}$Pa & 32 & 5.32 $\times$ $10^{-2}$ & 6.91 $\times$ $10^{-4}$ \\
& $^{238}$U & $^{214}$Pb & 32 & 8.09 $\times$ $10^{-3}$ & 0 \\
& $^{238}$U & $^{214}$Bi & 32 & 2.2 & 0.14 \\
& $^{238}$U & $^{210}$Tl & 6.4 $\times$ $10^{-3}$ & 5.97 $\times$ $10^{-4}$ & 5.50 $\times$ $10^{-5}$ \\
& $^{238}$U & $^{210}$Bi & 32 & 6.22 $\times$ $10^{-3}$ & 0 \\
& $^{232}$Th & $^{228}$Ac & 25 & 0.419 & 1.34 $\times$ $10^{-3}$ \\
& $^{232}$Th & $^{212}$Pb & 25 & 8.00 $\times$ $10^{-5}$ & 0 \\
& $^{232}$Th & $^{212}$Bi & 16 & 9.04 $\times$ $10^{-2}$ & 2.66 $\times$ $10^{-3}$ \\
& $^{232}$Th & $^{208}$Tl & 9 & 1.76 & 0.314 \\
& $^{40}$K & $^{40}$K & 97 & 0.877 & 2.84 $\times$ $10^{-2}$\\ \\
\textbf{Liner} & $^{238}$U & $^{234}$Pa & 2.91 & 0.931 & 0.130 \\
& $^{238}$U & $^{214}$Pb & 2.91 & 8.13 $\times$ $10^{-2}$ & 2.87 $\times$ $10^{-4}$ \\
& $^{238}$U & $^{214}$Bi & 2.91 & 1.38 & 0.230\\
& $^{238}$U & $^{210}$Tl & 5.82 $\times$ $10^{-4}$ & 3.96 $\times$ $10^{-4}$ & 1.29 $\times$ $10^{-4}$ \\
& $^{238}$U & $^{210}$Bi & 2.91 & 0.310 & 6.21 $\times$ $10^{-3}$ \\
& $^{232}$Th & $^{228}$Ac & 0.199 & 4.57 $\times$ $10^{-2}$ & 1.97 $\times$ $10^{-3}$ \\
& $^{232}$Th & $^{212}$Pb & 0.199 & 2.44 $\times$ $10^{-4}$ & 0 \\
& $^{232}$Th & $^{212}$Bi & 0.127 & 4.08 $\times$ $10^{-2}$ & 5.34 $\times$ $10^{-3}$ \\
& $^{232}$Th & $^{208}$Tl & 7.16 $\times$ $10^{-4}$ & 5.04 $\times$ $10^{-2}$  & 1.97 $\times$ $10^{-2}$ \\
& $^{40}$K & $^{40}$K & 5.08 & 1.25 & 6.33 $\times$ $10^{-2}$ \\ 
     
\bottomrule
\end{tabular}
\end{table}

\begin{table}[t!]
\centering
\caption{Calculated activities and trigger rates for dominant decay chains for the PSUP, PMTs and encapsulation. Trigger rates are shown for individual isotopes and decay products. The last column shows the rate for $>3$ PMTs registering a hit in a 100~ns time window. This the default trigger condition. 
}
\label{tab:backgroundtriggerspmt}
\begin{tabular}{c c c c c c c}
\toprule
& & & \textbf{Activity} & \textbf{Detection} & \textbf{Detection rate}\\
\textbf{Component} & \textbf{Decay chain} & \textbf{Isotope} & \textbf{rate [Bq]} & \textbf{rate [Hz]} & \textbf{(hits>3) [Hz]} \\
\midrule
\textbf{PSUP} & $^{238}$U  & $^{234}$Pa & 6.56 & 1.02 & 3.23 $\times 10^{-2}$ \\
& $^{238}$U  & $^{214}$Pb & 6.56     & 9.01 $\times 10^{-2}$ & 8.49 $\times$ $10^{-5}$ \\
& $^{238}$U  & $^{214}$Bi & 6.56     & 2.06     & 0.211    \\
& $^{238}$U  & $^{210}$Tl & 1.31 $\times 10^{-3}$ & 6.19 $\times$ $10^{-4}$ & 1.00 $\times$ $10^{-4}$ \\
& $^{238}$U  & $^{210}$Bi & 6.56     & 0.309    & 1.42 $\times$ $10^{-3}$ \\
& $^{232}$Th & $^{228}$Ac & 0.564    & 7.27 $\times$ $10^{-2}$ & 8.35 $\times 10^{-4}$ \\
& $^{232}$Th & $^{212}$Pb & 0.564    & 4.49 $\times$ $10^{-4}$ & 0 \\
& $^{232}$Th & $^{212}$Bi & 0.361  & 5.76 $\times$ $10^{-2}$ & 1.97 $\times$ $10^{-3}$ \\
& $^{232}$Th & $^{208}$Tl & 0.203  & 0.114    & 3.57 $\times$ $10^{-2}$ \\
& $^{40}$K   & $^{40}$K   & 4.92     & 0.599    & 1.49 $\times$ $10^{-2}$ \\ \\

\textbf{PMT} & $^{238}$U & $^{234}$Pa & 155 & 91.7 & 8.15 \\
& $^{238}$U & $^{214}$Pb & 155 & 22.5 & 4.09 $\times$ $10^{-2}$ \\
& $^{238}$U & $^{214}$Bi & 155 & 105 & 18.1\\
& $^{238}$U & $^{210}$Tl & 3.1 $\times$ $10^{-2}$ & 2.62 $\times$ $10^{-2}$ & 9.11 $\times$ $10^{-3}$ \\
& $^{238}$U & $^{210}$Bi & 155 & 49.6 & 0.571 \\
& $^{232}$Th & $^{228}$Ac & 101 & 45.9 & 1.59 \\
& $^{232}$Th & $^{212}$Pb & 101 & 1.95 & 4.85 $\times$ $10^{-4}$ \\
& $^{232}$Th & $^{212}$Bi & 64.6 & 37.7 & 3.24 \\
& $^{232}$Th & $^{208}$Tl & 36.4 & 29.5  & 13.4 \\
& $^{40}$K & $^{40}$K & 10.7 & 5.62 & 0.188 \\ \\
\textbf{Encapsulation} & $^{238}$U  & $^{234}$Pa & 26.5     & 7.33     & 0.716    \\
& $^{238}$U  & $^{214}$Pb & 26.5     & 1.16     & 3.18 $\times$ $10^{-3}$ \\
& $^{238}$U  & $^{214}$Bi & 26.5     & 11.5     & 1.76     \\
& $^{238}$U  & $^{210}$Tl & 5.30 $\times$ $10^{-3}$ & 3.10 $\times$ $10^{-3}$ & 9.07 $\times$ $10^{-4}$ \\
& $^{238}$U  & $^{210}$Bi & 26.5     & 3.16     & 4.50 $\times$ $10^{-2}$ \\
& $^{232}$Th & $^{228}$Ac & 9.87     & 2.25     & 7.08 $\times$ $10^{-2}$ \\
& $^{232}$Th & $^{212}$Pb & 9.87     & 3.84 $\times$ $10^{-2}$ & 0        \\
& $^{232}$Th & $^{212}$Bi & 6.32     & 1.76     & 0.162    \\
& $^{232}$Th & $^{208}$Tl & 3.55     & 2.31     & 0.899    \\
& $^{40}$K   & $^{40}$K   & 87.8     & 20.9     & 0.835   \\   
\bottomrule
\end{tabular}
\end{table}

Simulation studies are made to estimate the background rates from the assay data (Section~\ref{sec:clean}). These are carried out within the RATPAC-Two \cite{ratpac2} framework, which is based on GEANT4 \cite{GEANT4:2002zbu,Allison:2016lfl}, CLHEP \cite{CLHEP}, GLG4sim \cite{GLG4sim}, and ROOT \cite{ROOT}. Each object used to construct \BUTTON contains radioactivity from multiple decay chains, each chain containing multiple decays that emit charged particles that can interact to produce photons. Therefore, all decay processes of interest must be run in order to determine a realistic detector background rate. The setup of the background simulations is handled by the COBRAA package described in~\cite{Kneale:2022vpw} which generates a RAT-PAC macro for each type of event generation required. The results of these studies are summarized in Tables~\ref{tab:backgroundtriggerstank} and \ref{tab:backgroundtriggerspmt}. For the $^{238}$U chain, the main contribution comes from $^{214}$Bi, while for the $^{232}$Th chain, $^{208}$Tl dominates. The expected background trigger rate from these studies is $\sim\!120\,\mathrm{Hz}$, dominated by the contribution from the rock of the cavern.

\section{Summary}
\label{sec:summary}
The \BUTTON detector will test the performance of hybrid neutrino detection technology in the underground environment of the Boulby Laboratory. Detector installation is complete, and the experiment is now ready to begin data collection in Autumn 2025. The first fills will determine the baseline detector performance with pure water. Subsequently, fills with WbLS and Gd doping will be made. This program will provide important input to the international effort to develop and evaluate this new technology in an ultra-low background environment. Beyond the baseline goals, the flexible design of the experiment means future fills with alternative media are possible. The information gained on cosmogenic and radioactive background levels in Boulby Mine will be invaluable for future experiments at BUL.

\acknowledgments
 The Boulby Underground Laboratory is funded and operated by the Science and Technologies Facilities Council (STFC) operating under United Kingdom Research and Innovation. The laboratory is located in an active polyhalite mine near Whitby on the north-east coast of England, operated by ICL Boulby. The authors thank both the STFC Boulby staff and the management and staff of ICL. Without their cooperation and support, the installation of \BUTTON would not have been possible. This work was supported in the UK by the Science and Technology Facilities Council (Grant numbers: ST/S006400/1, ST/S006419/1, ST/S006524/1, ST/T005319/1, ST/V002341/1, ST/Y003373/1, ST/Y003500/1, ST/Y003462/1, ST/Y003411/1) and by the Atomic Weapons Establishment (AWE), as contracted by the Ministry of Defence. The work of Ben Richards is supported by UK Research and Innovation (UKRI) grant MR/Y034082/1. 

Work was performed under the auspices of the U.S. Department of Energy by Lawrence Livermore National Laboratory under contract DE-AC52-07NA27344. LLNL-JRNL-2011933. The work was supported in part by the U.S. Department of Energy National Nuclear Security Administration Office of Defence Nuclear Nonproliferation R\&D through the Consortium for Monitoring, Technology, and Verification under award number DE-NA0003920. The work carried out at Lawrence Berkeley National Laboratory was carried out under the auspices of the US Department of Energy under the contract DE-AC02-05CH11231. The project was funded by the U.S. Department of Energy, National Nuclear Security Administration, Office of Defense Nuclear Nonproliferation  Research and Development (DNN R\&D). This material is based upon work supported by the U.S. Department of Energy, Office of Science, Office of High Energy Physics, under Award Number  DE-SC0018974. 

\noindent{}For the purpose of open access, the author has applied a Creative Commons Attribution (CC BY) licence to any Author Accepted Manuscript version arising from this submission.

\providecommand{\href}[2]{#2}\begingroup\raggedright\endgroup

\end{document}